\newcommand{\Tr}{\mathop{\mathrm{Tr}}\nolimits}

\documentclass[prl,twocolumn,showpacs,superscriptaddress]{revtex4}

\usepackage{amsfonts,amssymb,bm,graphicx}

\begin{document}

\title{Maximally polarized states for quantum light fields}

\author{Luis L. S\'anchez-Soto}
\affiliation{Departamento de \'{O}ptica, Facultad de F\'{\i}sica,
Universidad Complutense, 28040~Madrid, Spain}

\author{Eulogio C. Yustas}
\affiliation{Departamento de \'{O}ptica, Facultad de F\'{\i}sica,
Universidad Complutense, 28040~Madrid, Spain}

\author{Gunnar Bj\"{o}rk}
\affiliation{School of Information and Communication Technology,
Royal Institute of Technology (KTH), Electrum 229, SE-164 40 Kista,
Sweden}

\author{Andrei B. Klimov}
\affiliation{Departamento de F\'{\i}sica,
Universidad de Guadalajara, 44420~Guadalajara,
Jalisco, Mexico}

\date{\today}

\date{\today}

\begin{abstract}
degree of polarization of a quantum state can be
defined as its Hilbert-Schmidt distance to the set of
unpolarized states. We demonstrate that the states
optimizing this degree for a fixed average number of
photons $\bar{N}$ present a fairly symmetric, parabolic
photon statistics, with a variance scaling as $\bar{N}^2$.
Although no standard optical process yields such a
statistics, we show that, to an excellent approximation,
a highly squeezed vacuum can be considered as maximally
polarized.
\end{abstract}

\pacs{42.50.Dv, 03.65.Yz, 03.65.Ca, 42.25.Ja}

\maketitle

A number of key concepts in quantum optics can be
concisely quantified in terms of distance measures.
The notions of nonclassicality~\cite{nonclass},
entanglement~\cite{entang}, information~\cite{accinf},
localization~\cite{local}, or polarization~\cite{pol},
to cite only a few relevant examples, have been
systematically formulated within this framework.
The rationale behind this is quite clear: once
we have identified a convex set with the desired
physical properties (classicality, separability, etc.),
the distance determines the distinguishability of a
state with respect to that set~\cite{Wootters81}.
Apart from its conceptual simplicity, this procedure
avoids many undesired problems that can arise in more
standard approaches.

Irrespective of our particular choice for the distance,
a natural question emerges: what states maximize the
corresponding measure. A good deal of effort has been
devoted to characterize maximally nonclassical or
entangled states. However, as far as we know,
maximally polarized states have been not considered
thus far, except for some trivial cases. It is precisely
the purpose of this Letter to fill this gap, providing a
complete description of such states, as well as
feasible experimental schemes for their generation.

Let us start by briefly recalling some basic concepts
about quantum polarization. We assume a monochromatic
plane wave propagating in the $z$ direction, whose electric
field lies in the $xy$ plane. Under these conditions,
the field can be fully represented by two complex
amplitude operators, denoted by $\hat{a}_H$ and
$\hat{a}_V$ when using the basis of linear (horizontal
and vertical) polarizations. They obey the standard
bosonic commutation relations $[\hat{a}_j, \hat{a}_k^\dagger ]
= \delta_{jk}$, with $j, k \in \{H, V \}$. The Stokes
operators are then introduced as the quantum counterparts
of the classical variables~\cite{Stokes}, namely
\begin{eqnarray}
\label{Stokop}
& \hat{S}_0 = \hat{a}^\dagger_H \hat{a}_H +
\hat{a}^\dagger_V \hat{a}_V \, ,
\qquad
\hat{S}_1 = \hat{a}^\dagger_H \hat{a}_V +
\hat{a}^\dagger_V \hat{a}_H \, , &
\nonumber \\
& & \\
& \hat{S}_2 = i ( \hat{a}_H \hat{a}^\dagger_V -
\hat{a}^\dagger_H \hat{a}_V ) \, ,
\qquad
\hat{S}_3 = \hat{a}^\dagger_H \hat{a}_H -
\hat{a}^\dagger_V \hat{a}_V \, , &
\nonumber
\end{eqnarray}
and their mean values are precisely the Stokes
parameters $(\langle \hat{S}_0 \rangle, \langle
\hat{\mathbf{S}} \rangle )$, where $\hat{\mathbf{S}}^T =
(\hat{S}_1, \hat{S}_2, \hat{S}_3)$ and the superscript
$T$ indicates the transpose. One immediately gets
that the Stokes operators satisfy the commutation
relations of the algebra su(2): $[ \hat{S}_1, \hat{S}_2]
= 2 i \hat{S}_3$, and cyclic permutations. Moreover,
since $[\hat{\mathbf{S}}, \hat{S}_0] = 0$, each energy
manifold (with a fixed number of photons) can be treated
separately. To bring out this point more clearly, it
is advantageous to relabel the standard two-mode Fock
basis $|n \rangle_H \otimes |m \rangle_V$ in the form
\begin{equation}
\label{invsub}
| N, k \rangle = | k \rangle_H \otimes
|N - k \rangle_V ,
\qquad
k = 0, 1, \ldots,  N .
\end{equation}
These states span an invariant subspace of dimension $N+1$,
and the operators $\hat{\mathbf{S}}$ act therein as an
angular momentum $N/2$.

Any (linear) polarization transformation is generated
by the Stokes operators (\ref{Stokop}). However, $\hat{S}_0$
induces only a common phase shift that does not change
the polarization state and can thus be omitted. Therefore,
we restrict ourselves to the SU(2) transformations,
generated by $\mathbf{S}$. Since $\hat{S}_1$ is related to
$\hat{S}_2$ and $\hat{S}_3$ by the commutation relations,
only these two generators suffice. It is well known that
$\hat{S}_2$ generates rotations around the direction of
propagation, whereas $\hat{S}_3$ represents differential
phase shifts between the modes. It follows then that any
polarization transformation can be realized with linear
optics: phase plates and rotators.

Although a precise defintion of polarized light at the
quantum level may be controversial~\cite{Klyshko96}, there
is a wide consensus~\cite{unpol} in viewing unpolarized
states as the only ones that remain invariant under any
polarization transformation. It turns out that the density
operator of these states can be written
as
\begin{equation}
\label{denunpol}
\hat{\sigma} = \sum_{N=0}^\infty p_N
\left ( \frac{1}{N+1} \ |N, k \rangle
\langle N, k | \right )  \, ,
\end{equation}
where $p_N$ is the two-mode photon-number distribution. Therefore,
$\hat{\sigma}$ appears as diagonal in every subspace, with
coefficients that guarantee the unit trace condition.

According to our previous discussion, it seems natural
to quantify the degree of polarization of a state
represented by the density operator $\hat{\varrho}$ as
$\mathbb{P} (\hat{\varrho}) \propto \inf_{\hat{\sigma} \in
\mathcal{U}} D(\hat{\varrho} , \hat{\sigma} )$, where
$\mathcal{U}$ denotes the convex set of unpolarized
states of the form (\ref{denunpol}) and $D(\hat{\varrho},
\hat{\sigma} )$ is any measure of distance between the two
density matrices $\hat{\varrho}$ and $\hat{\sigma}$. The
degree $\mathbb{P} (\hat{\varrho} )$ must be normalized to
the unity and satisfy some requirements motivated by both
physical and mathematical concerns~\cite{entang}. There are
numerous nontrivial choices for $D(\hat{\varrho} ,
\hat{\sigma})$: we shall consider here the Hilbert-Schmidt
metric $D_{\mathrm{HS}} (\hat{\varrho}, \hat{\sigma}) =
|| \hat{\varrho} - \hat{\sigma} ||_2^2 =   \Tr [(
\hat{\varrho} - \hat{\sigma})^2 ]$ because is the
simplest one for computational purposes and allows
us to get an analytic form of the degree of polarization
\begin{equation}
\label{PHS}
\mathbb{P} (\hat{\varrho} ) =
\Tr ( \hat{\varrho}^2 ) -
\sum_{N=0}^\infty \frac{p_N^2}{N + 1} \, ,
\end{equation}
which is determined by the purity $0  < \Tr ( \hat{\varrho}^2 )
\leq 1$ and the photon-number distribution $p_N$.

It is clear from  Eq.~(\ref{PHS}) that for the states living
in the manifold with exactly $N$ photons, the optimum is
reached for pure states [for which $ \Tr ( \hat{\varrho}^2 ) = 1$].
In fact, all such pure states have the same degree of polarization:
\begin{equation}
\label{PHSNs} \mathbb{P}_N = \frac{N}{N + 1} \sim  1 -
\frac{1}{N} \, ,
\end{equation}
where the last expression, showing the typical scaling $N^{-1}$,
holds when $N \gg 1$. In particular, the important SU(2)
coherent states on the Poincar\'e sphere~\cite{Per86}
\begin{equation}
|N, \theta, \phi \rangle  =
\sum_{k=0}^N
C_k (\theta, \phi) \  | N, k \rangle \, ,
\end{equation}
with coefficients
\begin{equation}
C_k (\theta, \phi)  =
\left (
\begin{array}{c}
N \\
k
\end{array}
\right )^{1/2}
\left ( \sin \frac{\theta}{2} \right)^{N-k}
\left ( \cos \frac{\theta}{2} \right)^k
e^{-i k \phi}
\end{equation}
are such $N$-photon states.

However, this scaling law $N^{-1}$ can be easily surpassed.
Perhaps the simplest example is when both modes are in
(quadrature) coherent states,  we denote by $| \alpha_H,
\alpha_V \rangle$. By reparametrizing the amplitudes
as $\alpha_H = e^{-i\phi /2} \sqrt{\bar{N}} \sin ( \theta/2)$
and $\alpha_V = e^{i\phi /2} \sqrt{\bar{N}} \cos ( \theta/2)$,
we can express them as
\begin{equation}
\label{qucoh}
| \alpha_H, \alpha_V \rangle =
\sum_{N=0}^\infty
e^{- \bar{N}/2} \frac{ \bar{N}^{N/2}}{\sqrt{N!}} \,
|N, \theta, \phi \rangle \, ,
\end{equation}
where $\bar{N} = |\alpha_H|^2 + |\alpha_V|^2$
is the average number of photons. Since $p_N$ is
a Poissonian with mean value $\bar{N}$, one
can perform the sum in (\ref{PHS}), with the
result
\begin{equation}
\mathbb{P}_{\mathrm{coh}} = 1 -
\frac{I_1 (2 \bar{N})}
{\bar{N}} e^{-2 \bar{N}} \sim  1-
\frac{1}{2\sqrt{\pi} \bar{N}^{3/2}} \, .
\end{equation}
where $I_1 (z)$ is the modified Bessel function and
the last equality is valid for $\bar{N} \gg 1$.

In consequence, we are led to find optimum states
for a fixed average number of photons $\bar{N}$.
Obtaining the whole optimum distribution $p_N$ in
(\ref{PHS}) is exceedingly difficult, since it
involves optimizing over an infinite number of
variables. Our strategy to attack this problem
is to truncate the Hilbert space and consider
only photon numbers up to some value $D$, where
we take the limit $D \rightarrow \infty$ at the end.
In this truncated space, we need to find the states
that maximize (\ref{PHS}) with the constraints
\begin{equation}
p_N \ge 0 , \qquad
\sum_{N=0}^D p_N = 1 , \qquad
\sum_{N=0}^D  N \ p_N = \bar{N} .
\end{equation}
It is clear that the optimum must be again pure states.
If we introduce the notations $p^T =  (p_0, p_1,
\ldots, p_D)$ and $ H =  2\  \mathrm{diag}
[1, 1/2, \ldots, 1/(D+1) ]$, the task can be thus
recast as
\begin{eqnarray}
\label{qp}
\text{minimize} & &
\frac{1}{2} p^T  H  p \nonumber \\
& & \\
\text{subject to} & &
A p = b  , \nonumber \\
& & p \ge 0 , \nonumber
\end{eqnarray}
where
\begin{equation}
b =
\left (
\begin{array}{c}
1 \\
\bar{N} \\
\end{array}
\right ) ,
\qquad
A =
\left (
\begin{array}{ccccc}
1 & 1 & 1 & \cdots & 1 \\
0 & 1 & 2 & \cdots & D  \\
\end{array}
\right ) .
\end{equation}
We deal then with a quadratic programming problem that,
in addition, is convex, because  $H$ is positive
definite~\cite{Boyd04}. The optimum point exists
and it is unique: in fact, there are numerous
algorithms that compute this optimum in a quite
efficient manner. Alternatively, we may try to determine
it analytically  by incorporating the constraints by
the method of Lagrange multipliers. The functional to
be minimized is
\begin{equation}
\mathcal{L} (p, \lambda) =
\frac{1}{2} p^T H p -
\lambda^T (A p -b) \, .
\end{equation}
The first-order optimality conditions $ \nabla
\mathcal{L} (p, \lambda) = 0$ together with the
initial equality constraint, give the system of
linear equations
\begin{equation}
\left (
\begin{array}{cc}
H & - A^T \\
A & 0
\end{array}
\right )
\left (
\begin{array}{c}
p \\
\lambda
\end{array}
\right )
 =
\left (
\begin{array}{c}
0  \\
b
\end{array}
\right )  \, ,
\end{equation}
whose formal solution is
\begin{equation}
\label{csol}
\lambda =  (A H^{-1} A^T)^{-1} b \, ,
\qquad
p =   H^{-1} A^T  \lambda \, .
\end{equation}

\begin{figure}
\includegraphics[width=0.85\columnwidth]{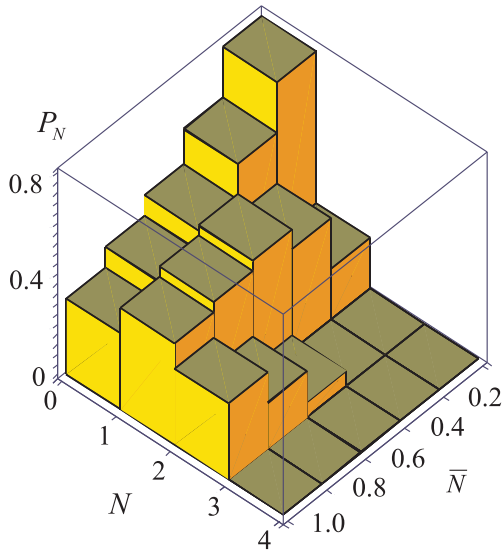}
\caption{Optimum distribution  $p_N$, obtained by
solving numerically the quadratic program (\ref{qp}),
plotted as a function of the average number of photons
$\bar{N}$ and $N$. We have taken $\bar{N}$ running
from 0.2 to 1 and the dimension of the space $D = 4$.}
\end{figure}

Before working out the analytical form of (\ref{csol}),
in Fig.~1 we have plotted the numerical solution of the
quadratic program (\ref{qp}) for some values in
$0 \le \bar{N} \le 1$, using the MINQ code implemented
in Matlab. The number of nonzero components of $p_N$ is
$[2 \bar{N} + 1]$, where the brackets denote the integer
part. The distribution presents a clear skewness and
one can check that it can be well fitted to a
Poisson distribution, which in physical terms  means
that, in this range, a  quadrature coherent state
$| \alpha_H, \alpha_V \rangle$ can be considered as
optimum.  To better assess this behavior, we have
calculated the associated Mandel $Q$
parameter~\cite{Mandel95}
\begin{equation}
Q =  \frac{\langle (\Delta \hat{N} )^2 \rangle}
{\langle \hat{N} \rangle} - 1 \, ,
\end{equation}
where $\langle (\Delta \hat{N} )^2 \rangle$ is the variance,
which is a standard measure of the deviation from the
Poisson statistics. In Fig.~2 we have represented $Q$
in terms if $\bar{N}$. As we can see, $Q$ increases
linearly with $\bar{N}$ and is zero only near
$\bar{N} \simeq 3$.

\begin{figure}
\includegraphics[width=0.85\columnwidth]{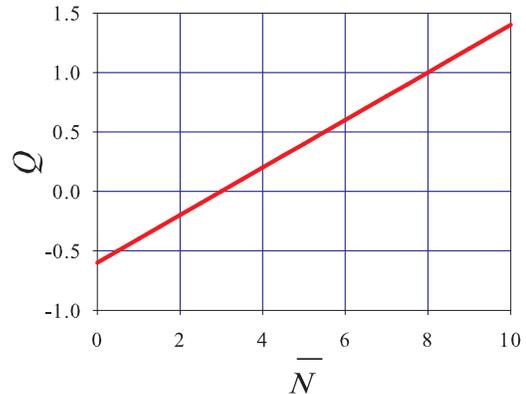}
\caption{Plot of the Mandel $Q$ parameter for the
optimum distribution $p_N$ obtained numerically from
(\ref{qp}) in terms of the average number of photons $\bar{N}$
of the state.}
\end{figure}

In Fig.~3 we have plotted the optimum distribution $p_N$
for different integer values of $\bar{N}$ running from 1
to 9. The truncation value has been chosen to be 25 in
all the cases, although it is sufficient to ensure,
for each value of $\bar{N}$, that $p_N \simeq 0$ for
$N > D$. Three distinctive features can be immediately
discerned: the solutions are symmetric around $\bar{N}$,
they are parabolic, and extend in a range from 0 to ç
$2 \bar{N}$. The two first facts are in agreement with
the symmetry properties of the original problem (\ref{qp}).
The third one means a variance that scales as $\bar{N}^2$,
at difference of what happens for standard coherent
optical processes presenting a variance linear with
$\bar{N}$ (as for, e.g., in Poissonian or Gaussian statistics).
In other words, the optimum states are extremely noisy
and fluctuating. When $\bar{N}$ is not integer (or semi-integer),
one can appreciate a small asymmetry that is less and
less noticeable as $\bar{N}$ increases.

\begin{figure}[b]
\includegraphics[width=0.85\columnwidth]{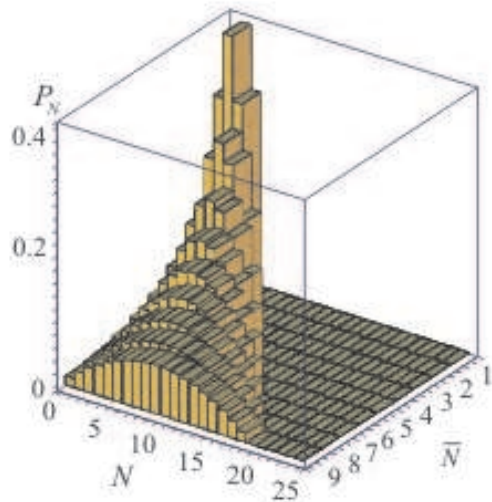}
\caption{Optimum distribution $p_N$ plotted as a
function of the average number of photons $\bar{N}$
and $N$. We have taken $\bar{N}$ to be a integer
running from 1 to 9 and the dimension of the space
$D = 25$.}
\end{figure}

We conclude that we can take the dimension $D$
to be $2 \bar{N}$. Given the very simple form
of $H$ and $A$, we can express the final
solution (\ref{csol}) in a closed analytic form:
\begin{equation}
\label{qs}
p_N = 3 \frac{(\bar{N} + 1)^2 - (N - \bar{N})^2}
{(2 \bar{N} +1 ) (\bar{N} + 1) ( 2 \bar{N} + 3)}
\simeq \frac{3}{2 \bar{N}}  \left ( \frac{N}{\bar{N}} -
\frac{1}{2} \frac{N^2}{\bar{N}^2} \right ) \, ,
\end{equation}
which is properly normalized and shows all the
aforementioned characteristics,  with a maximum
value of $p_{\bar{N}} \simeq  3/(4 \bar{N})$.
If we use $x = N/(2 \bar{N})$, which can be
considered as a quasicontinuous variable $0 \le x \le 1$,
we can convert (\ref{qs}) in $ p(x) = (3/\bar{N}) \ x (1-x)$,
which is the Beta distribution of parameters $(2, 2)$~\cite{Evans2000}.
For the solution (\ref{qs}), the corresponding degree
of polarization is
\begin{equation}
\label{pq2}
\mathbb{P}_{\mathrm{opt}} = 1 -
\frac{3}{(2 \bar{N} +1 ) ( 2 \bar{N} + 3)}
\sim 1 - \frac{3}{4 \bar{N}^2} \, .
\end{equation}
This provides a full characterization of the optimum states
we were looking for. However, their physical implementation
stands as a serious problem. The crucial issue for the
scaling in (\ref{pq2}) is the fact that distribution
variance is proportional to $\bar{N}^2$. It turns out
that, for the discrete uniparametric distributions
usually encountered in physics, this is distinctive
of the thermal (or geometric) distribution
\begin{equation}
\label{Ptwins}
p_N = \frac{1}{\bar{N} + 1} \left ( \frac{\bar{N}} {\bar{N} + 1 }
\right )^N  \, .
\end{equation}
But this is the photon statistics associated with the states
\begin{equation}
\label{maxentnn}
| \xi \rangle = \frac{1}{\cosh \xi} \,
\sum_{n=0}^\infty \tanh^n \xi \;
|n \rangle_H \otimes | n \rangle_V \, ,
\end{equation}
which are the twin beams generated in an optical parametric
amplifier with a vacuum-state input, with $\bar{N} = 2
\sinh^2 \xi$. The distribution (\ref{Ptwins}) presents a
skewness absent in the exact solution (\ref{qs}), but a
calculation of the state degree of polarization gives
\begin{equation}
\mathbb{P}_{\xi} \simeq 1 -
\frac{\ln (\bar{N}/2)}{\bar{N}^2} \, .
\end{equation}
Apparently, this is different from (\ref{pq2}), but as soon as
$\bar{N} \gg 1$ they both approach unity in essentially the
same way, which means that the (maximally entangled) squeezed
vacuum (\ref{maxentnn}) is very close to optimum when $\bar{N}
\gg 1$. In fact, one can calculate the degree of polarization
for (\ref{maxentnn}) using other metrics than the Hilbert-Schmidt.
For example, if one employs the Bures (or fidelity) distance,
a simple exercise shows that $\mathbb{P}^{\mathrm{Bures}}_\xi =
1 - 1/\bar{N}^2$, confirming again the fundamental scaling $\bar{N}^{-2}$.

Before ending, two important remarks seem in order. First, we
observe that in classical optics fully polarized fields have a
perfectly defined relative phase between $H$- and
$V$-polarized modes~\cite{Bro98}. Such a relation does not
necessarily hold in the quantum domain: while the quadrature
coherent states (\ref{qucoh}) have a sharp relative phase, the
twin-photon beams (\ref{maxentnn}) have an almost random relative
phase. Second, the maximally polarized states we have found have
a highly nonclassical behavior, even in the limit $\bar{N} \gg 1$,
which makes the classical limit of these polarized states a
touchy business.

We thank A. Felipe and M. Curty for helpful discussions.
This work was supported by the Swedish Foundation for
International Cooperation in Research and Higher Education
(STINT), the Swedish Foundation for Strategic Research (SSF),
the Swedish Research Council (VR), the CONACyT  grant PROMEP/103.5/04/1911,
and the Spanish Research Project FIS2005-0671.

\end{document}